\documentclass[sigconf]{acmart}

\usepackage{times}  
\usepackage{xcolor}
\renewcommand\footnotetextcopyrightpermission[1]{} 
\setcopyright{none}

\renewcommand\footnotetextcopyrightpermission[1]{}

\begin{document}

%%%%%%%%%%%% THIS IS WHERE WE PUT IN THE TITLE AND AUTHORS %%%%%%%%%%%%

\title{A Pluralist Approach to Democratizing Online Discourse}

\author{Jay Chen}
\affiliation{
 \institution{International Computer Science Institute}
}
\email{jchen@icsi.berkeley.edu}
 
\author{Barath Raghavan}
\affiliation{%
  \institution{University of Southern California}
}
\email{barath.raghavan@usc.edu}

\author{Paul Schmitt}
\affiliation{%
  \institution{Princeton}
}
\email{pschmitt@cs.princeton.edu}

\author{Tai Liu}
\affiliation{%
  \institution{NYU Abu Dhabi}
}
\email{tai.liu@nyu.edu}

\begin{abstract}

Online discourse takes place in corporate-controlled spaces thought by users to be public realms.
These platforms in name enable free speech but in practice implement varying degrees of censorship either by government edict or by uneven and unseen corporate policy. This kind of censorship has no countervailing accountability mechanism, and as such platform owners, moderators, and algorithms shape public discourse without recourse or transparency.

Systems research has explored approaches to decentralizing or democratizing Internet infrastructure for decades. In parallel, the Internet censorship literature is replete with efforts to measure and overcome online censorship. However, in the course of designing specialized open-source platforms and tools, projects generally neglect the needs of supportive but uninvolved `average' users. In this paper, we propose a pluralistic approach to democratizing online discourse that considers both the systems-related and user-facing issues as first-order design goals.

\end{abstract}
\maketitle

%%%%%%%%%%%%%  ABSTRACT GOES HERE %%%%%%%%%%%%%%

\section{Introduction}

\noindent \emph{``The best test of truth is the power of the thought to get itself accepted in the competition of the market.''}

\hfill Oliver Wendell Holmes\\

\noindent \emph{``In order to maintain a tolerant society, the society must be intolerant of intolerance.''}

\hfill Karl Popper\\

\noindent \emph{``We are creating a world that all may enter without privilege or prejudice accorded by race, economic power, military force, or station of birth. We are creating a world where anyone, anywhere may express his or her beliefs, no matter how singular, without fear of being coerced into silence or conformity. Your legal concepts of property, expression, identity, movement, and context do not apply to us. They are all based on matter, and there is no matter here.''}

\hfill John Perry Barlow
\newline

%freedom of speech, centralization creates bottleneck
Freedom of speech is key to democracy. A free and open Internet is key to modern global society. Yet how does one reconcile the perspectives above, which speak to fundamental issues and are at once in agreement and at odds?
Throughout the 20th century the greatest threat to free speech was that of \emph{government censorship}. Indeed, this persists under autocracies around the world (e.g., China, Iran) but it has fused with \emph{corporate censorship}, with autocracies using corporate partners to implement censorship, or corporate moderators unevenly removing unprofitable or unpalatable speech.

The character of discourse online is not only a matter of policy or culture, but also one of \emph{systems}: the systems that provide or prevent the exercise of free speech. The wide reach and accessibility of the early Internet enabled flourishing free speech, but centralization into a handful of platforms (e.g., Facebook, Twitter, Weibo, etc.) has surrendered control over online speech.
%gov't censorship and defining corporate censorship
Social media platforms provide users a `free' service with convenience, usability, and powerful network effects. Users treat these platforms as public realms, but they are private spaces that can unilaterally decide the content they serve~\cite{facebook_remove_post,twitter_ghost_delete}. These corporations use one-sided terms of service to justify the removal of content and to provide arbitrary exemptions, typically without explanation~\cite{moderation_difficulties, facebook_jail,facebook_climate}. This lack of transparency and accountability raises the question: why should corporations be responsible, de jure (e.g., in China) or de facto (e.g., in the US), for adjudicating speech at all?

%law, recent politics/news
In the US, Section 230 of the Communications Decency Act~\cite{section_230} effectively shields online platforms from responsibility for content posted by a third party. However, because social media platforms have elected to apply their own moderation policies to restrict speech (e.g., harassment or hate speech), they have successfully interjected themselves into a difficult position. Recent disputes regarding how to handle controversial political speech on such platforms have resulted in politicians simultaneously criticizing platforms for doing both too little and too much policing~\cite{twitter_facebook_moderation}. Indeed, recent responses by Facebook and Twitter have differed significantly~\cite{facebook_trump_posts, twitter_trump_posts}, which further demonstrates how ill-equipped corporations are in arbitrating such matters~\cite{black_zebra}. At the time of writing, proposals to weaken Section 230 further complicate the problem, because if such amendments are made, corporations would then have to assume both the responsibility of policing content and the risk of legal liability~\cite{section_230_scraped}.

%moderation and algorithmic bias
Legal systems have traditionally recognized that freedom of speech should be limited when it conflicts with other freedoms~\cite{freespeechexceptions, defamation1, defamation2, sedition, libel}. Spam, scams, and other abuses are typically removed because they are nearly universally unwelcome. However, moderation on the basis of political or ideological grounds is notoriously difficult because there is so much content to filter and because different people and communities have their own norms and expectations~\cite{facebook_moderators, moderation_difficulties}. Today, platform operators generally delegate moderation to individuals who manage the problem at smaller scales or within particular communities. For example, Facebook has approximately 15,000 community managers in the US who review flagged content~\cite{facebook_moderators}.
%moderator censorship, platform bias
While moderators can let their personal views bias their decisions~\cite{human_moderator_probs}, the bigger problem when it comes to political speech is that companies can unevenly apply their own policies in favor of profitable content~\cite{platformbias}.

In response, some have advocated the creation of so-called free-speech platforms such as Gab, 4chan, and 8chan, which seem to always yield high concentrations of hate speech and are notorious for their racism, misogyny, and homophobia~\cite{zannettou2018gab}. These and other recent historical examples suggest that moderation is critical for preserving an environment that is not filled with large amounts of gore and hate speech~\cite{facebook_moderators}. 

In this spectrum of options we have seen the promise and pitfalls of the approaches that one might consider aligned with Holmes (e.g., free discourse with an expectation of high-minded debate), Popper (e.g., uneven regulation, either by governments or corporations, to ensure preservation of a relatively open sphere), and Barlow (e.g., a free for all). Our observation is that there is a need to balance free speech with moderation and so we introduce another option:\\[0.5ex]
\emph{``[D]emocracy is the worst form of Government except for all those other forms that have been tried from time to time.''}

\hfill Winston Churchill\\[0.5ex]

This democratic political perspective lays the philosophical foundation for our paper and is aligned with notions of free expression and the marketplace of ideas common in Western democracies. It also recognizes that a truly free Internet allows for the possibility of undesirable pockets of content, but that they are not desired by the vast majority of people. Based on this premise, we define two objectives that communications platforms should ideally achieve, sketch a pluralist architecture for realizing these goals, and discuss the opportunities and challenges that result from our design.

\section{Goals} \label{sec:timbre_design}

The early Internet consisted of many autonomous providers with little centralized control of services or infrastructure. Today’s Internet, while physically distributed, consists of a few popular services whose authority is held in the hands of corporations. As observed by Liu et al.~\cite{liu2017barriers}:

\emph{``Changes in the Internet’s structure have taken place along two orthogonal axes over the past few decades, yet they are often conflated. The first axis concerns physical distribution—centralized vs. decentralized -- whether the physical resources being accessed for some service are located at a single machine (at one extreme) or dispersed across many machines all over the planet (at the other). The second axis concerns control -— whether the control over the service and the machines providing a service is held by a few versus spread across many individuals or organizations. The Internet of today is quite different from that of a few decades past along both axes: it has gone from partially centralized and democratic to distributed and feudal.''}

In this paper we are explicitly focused on the decentralization of control rather than that of physical resources, and we use the term ``democratization'' interchangeably with that notion of decentralization. Numerous systems have been developed in academia, industry, and as open source projects in attempts to re-democratize the Internet~\cite{ali2016blockstack, redecentralize, solidmit, namecoin, gnusocial, emercoin, mastodon, friendica, identica, riot, zeronet, benet2014ipfs, beaker, maidsafe, nextcloud, ring}. The core problems that these projects tackle predominantly focus on basic Internet infrastructure such as: naming, communication, data storage, serverless web applications, and digital currencies.
Rather than focusing purely on systems goals such as scalability, security, and efficiency, we include two meta-level goals:

\vspace{1mm}
\noindent
\textbf{Democratize control over content:} Because today's social media companies own the infrastructure, they possess inordinate power over online content. They control what content may be posted, whether it remains stored, how it is discovered by users, whether it is presented to users, and how it is presented to users. Democratizing control over content (i.e., spreading out power) counteracts digital authoritarianism by definition. This goal is not new and broadly mirrors the motivations behind Tim Berners-Lee's Solid~\cite{solidmit} project and many others in the context of re-decentralizing the Web. However, we add a less explored corollary, of considering how to prevent re-centralization.

\vspace{1mm}
\noindent
\textbf{Preserve moderated user experience:} Online media companies like Facebook hire armies of moderators to filter out content that is legitimately illegal or undesirable by users~\cite{facebook_moderators}. Our goal is to allow free speech while preserving the user experience of moderated services.
This goal is typically considered out of scope in systems and relegated to user interface researchers or software developers. However, we argue that the moderation \emph{system} is distinct from the moderation \emph{interface} and that the content moderation system is as critical to social media network systems as conventional concerns of naming, storage, distribution, security, privacy, and performance.

\section{Design}

There are many possible ways to democratize social media. We take a clean slate approach that attends to the practical and potential paths to adoption. We borrow from existing ideas where possible to sketch out a general design that can be discussed and improved upon.

\subsection{Decoupling Content from Presentation}

To enable free speech, we need to democratize control over content so that anyone can post and read any content that they want. However, as we have discussed, free-for-all approaches often degenerate into chaos so it is critical to complement free speech with some form of moderation. Unfortunately, any non-trivial moderation scheme will censor or de-emphasize certain content. Free speech and moderation seem fundamentally at odds, but this apparent paradox can be partially resolved by observing that social media platforms could be split into two layers that address the corresponding concerns of content storage and presentation. In this manner, what data is on the system is decoupled from whether users will actually see it. Thus, the moderated user experience can be preserved while allowing free speech.\footnote{We discuss the issue of how to deal with unconscionable or illegal content (e.g., child porn) in Section~\ref{sec:discussion}.}

Decoupling storage from presentation is not a new idea. Existing techniques such as filtering content, blocking users, user preferences, and even search could be construed as all being variants of this approach since the underlying content still exists on the platform, but is simply not visible to a particular user at a particular time. We propose to make this decoupling explicit and push it to its logical extreme. We envision content storage and content presentation as wholly separate layers, each with open standards, protocols, and, potentially, legislation to support cross-service interoperability and data migration. In this environment, a user could, for example, have their social media data or posts stored on Amazon servers while having their `feed' served by Facebook.\footnote{This example is meant to be illustrative in terms of conceptualizing the technical feasibility as well as its pragmatic impracticality. Without parallel legislative measures, this arrangement is likely to be unstable since it would be more efficient and profitable to have a single entity provide both services while maintaining a monopoly on the user's data, which is the case today. We discuss more interesting examples enabled by our design in Section~\ref{sec:discussion}.}

\subsection{Democratizing Content and Moderation}

The complete democratization of social media platforms requires decentralization of infrastructure. Rather than the above user having their data hosted by Amazon, they could host it themselves, in collaboration with their social network, or in a hybrid configuration. The ubiquity of physical computing resources today makes this relatively trivial in terms of the amount of resources required and the technology required for decentralizing control over physical resources already exists~\cite{liu2017barriers}. Content distribution follows a similar line of reasoning and could be enabled through P2P protocols. 

Decentralizing presentation is less well explored. We conceive of presentation as being composed of two parts. The first part is moderation and the second is presentation itself. Content moderation is a relatively well-studied topic, but mostly in the narrow sense of the systems that exist in the wild rather than novel experimental schemes evaluated in-situ~\cite{platformbias, socialmediabias, partisansharing, echochambers}. For example, content moderation on Wikipedia has been exhaustively considered~\cite{barnstars}. Collaborative filtering was a hot topic for quite some time as well~\cite{collaborativefiltering}. Alternatively, content \emph{curation} (e.g., the Twitter retweet model or voting/ranking/sorting schemes), could be understood as the converse of moderation, where instead of deny lists blocking content, allow lists or ranking algorithms make interesting content more visible to users. Both content curation and ranking algorithms have been extensively studied~\cite{twittersurvey}. 

As with the decoupling of content from presentation, we propose to take democratizing presentation to its logical extreme so that content moderation is altered in two ways. 

First, we \textit{democratize moderation} so that any user can moderate any content. Moderators on today's platforms are beholden to the platform's operators. In contrast, with content and moderation decoupled, this no longer necessarily the case; any user is free to publish moderation streams. As is the case today, moderation could be performed by hand or automated~\cite{automoderator, cheng2015antisocial, chandrasekharan2019crossmod}. Interestingly, since moderation streams are data, moderation streams could be designed recursively to allow users to publish aggregated moderation streams or moderation streams of moderation streams.

Second, we \textit{democratize presentation}, eliminating the false notion of a single universal perspective or truth. Traditional conceptions of decentralized moderation focus on the decentralization of the task of moderation, but the eventual output is assumed to be singular and universal. For example, in Wikipedia a single article is available by default for all users. There is a belief in a collaboratively-edited universal truth, while any disagreements can be explored through the edit history for individual wiki pages. Similarly, in the case of Reddit, each forum has a set of moderators who define what is normatively acceptable content for everyone in the forum and the corporation has ultimate authority to censor content, individuals, and moderators themselves. We envision a pluralism of moderation streams that enable users to compare and decide which view(s) to adopt as their `interface' to the underlying multitude of `ground truth' content. Collaborative filtering tools such as blocktogether~\cite{blocktogether}, that allow users to subscribe to each other's block lists on Twitter are the closest to what we propose in terms of moderation, but our design is more transparent, open, and general system across platforms.

\subsection{Preserving User Experience}

%aggregators and analogous entities
By decoupling content and moderation, the state of online social media reverts to something akin to the early days of the Web, IRC, or bittorrent. The problems of content hosting, discovery, and aggregation, which originally motivated cloud providers, search engines, and torrent aggregators analogously necessitate a set of social media \emph{aggregators} that cache, index, and present content and their associated moderation feeds to users through convenient user interfaces.

%performance
Through caching and other standard optimization techniques, the performance of aggregators could be optimized to a point where the user experience is comparable to existing platforms in terms of performance and content. As an existence proof, Steemit~\cite{steemit} is a blockchain-based online social media network that is usable and performant. Although Steemit handles moderation differently from what we propose, its existence demonstrates that a P2P social media network can be performant and sufficiently usable at a reasonable scale (\textasciitilde1 million users), despite additionally running a distributed consensus protocol.

%usability
Beyond performance, we posit that distributed storage infrastructures are not in mainstream use by the `average' user because there is: a) no `killer app'; and b) insufficient incentive.
Centralized systems are attractive to users due to convenience, homogeneity, and cost~\cite{liu2017barriers}, but given comparable performance, interacting with aggregators and client-end apps could become indistinguishable from existing platforms for average users. Additional interface elements, interactions, and incentive schemes afforded by pluralism can be experimented with in the same way that new features and platforms are introduced today.

\subsection{Preventing Re-centralization}

The free and open source software (FOSS) concept, free software movement, open-source software movement, and open data initiatives all try to address the issue of closed systems. Despite these efforts, private platforms simply build their own network infrastructure, software stacks, and protocols to create walled gardens. Thus, even our clean slate `open' design requires additional measures to prevent the re-emergence of centralization. Beyond relying on legislation mandating open data or open protocols and standards, there are very few purely-technical solutions to this problem. While one potential path forward is to legislate, we introduce the concept of \emph{intentional impermanence} as a potential design resource for consideration.

Given the seemingly inevitable centralization brought on by economies of scale in a capitalistic society, intentional impermanence is intended to make it more difficult to `pin down' the platform for permanent capture by any party. In our design, intentional impermanence means making it easy for users to migrate to alternative infrastructures with little disruption. Impermanence and democratization are complementary. If, for example, a group of users does not like the users who moderate a forum, they can start their own moderation stream with very little overhead and no data loss. Alternatively, if a group of users discover that an underlying data storage provider has been compromised or co-opted in any way, they can switch to another provider with minimal service interruption. Designing for impermanence mitigates risk of hostile takeovers because it disincentivizes adversaries to make the attempt with little to gain.

\section{Architecture}

\begin{figure}[t]
\centering
\includegraphics[width=1\columnwidth]{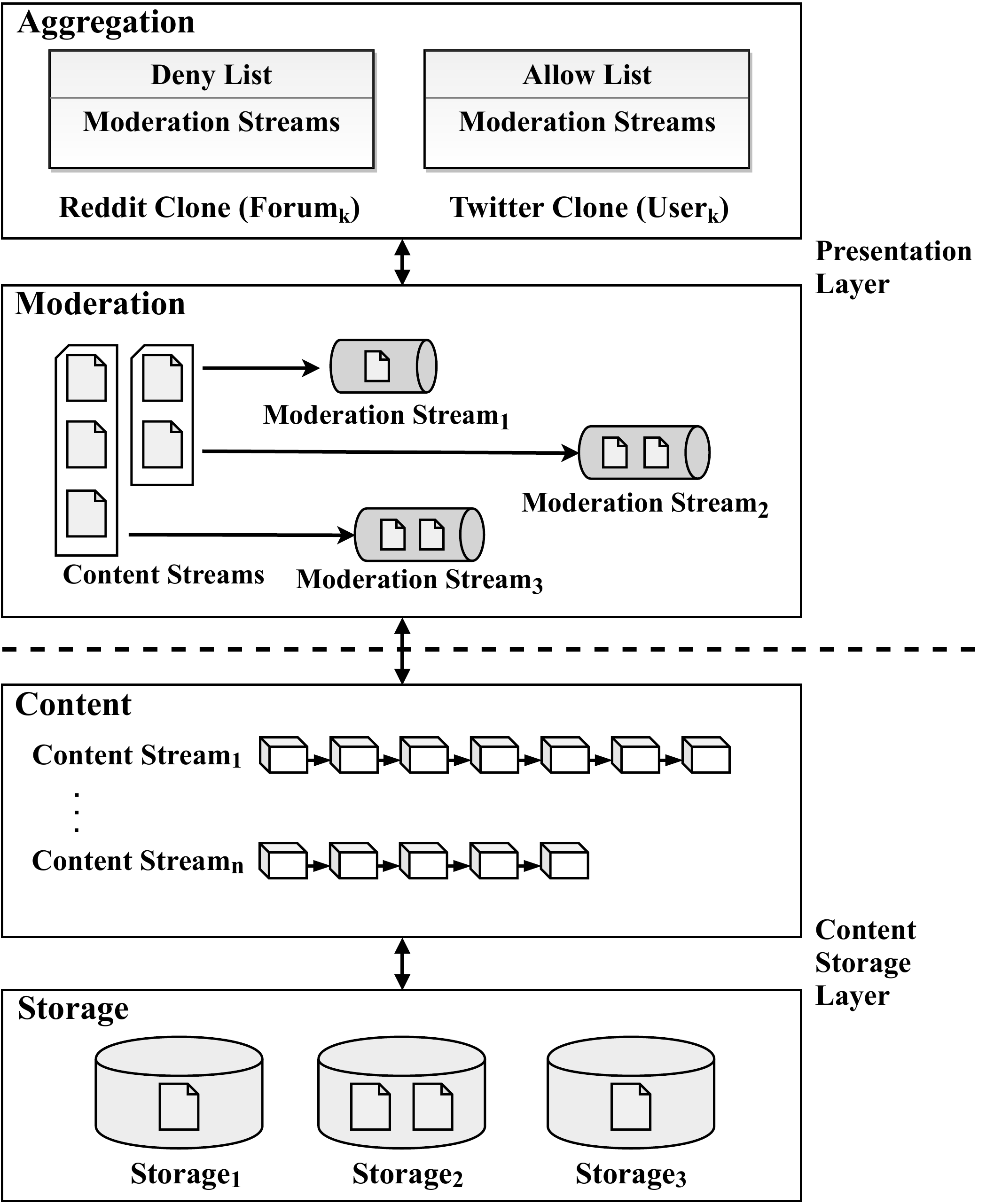}
\caption{Architecture layers and logical components.} 
\label{fig:arch}
\vskip -1.4em
\end{figure}

We propose an architecture that consists of four logical components: a content layer, a storage layer, a moderation layer, and an aggregation layer (Figure~\ref{fig:arch}). Here, we describe the components in the abstract while offering a bare-minimum means by which each piece could be implemented and leave the deliberation of an `optimal' solution for future work.

\vspace{1mm}
\noindent
%not sure whether to call them indexes or streams or what...
\textbf{Content Streams:} A \emph{content stream} is a data structure that maintains an index of all content (e.g., posts, tweets, etc.) published by a user or collaboratively by a set of users to support the implementation of forums, groups, or wikis. Content streams could be implemented in different ways, but minimally store a publisher-signed hash of the content being stored and a pointer to retrieve the content from storage. One trivial implementation of content streams could be in JSON format and stored in a local database. A more sophisticated implementation could be as a blockchain to incorporate potentially useful properties such as immutability, distributed consensus, or a native currency.

\vspace{1mm}
\noindent
\textbf{Storage:} Content streams index published content, but the content itself is stored in user-controlled physical storage that could include local disk, existing cloud-providers, or other distributed storage systems~\cite{benet2014ipfs}. As such, users may provide space to store and distribute other users' content for free or for a fee. Replication and distribution of content provides increased availability, reliability, and performance at costs determined by the market value of storage and bandwidth.

\vspace{1mm}
\noindent
\textbf{Moderation:} Rather than arbitrarily assigning special privileges to a small set of moderators, we allow anyone to become a moderator of any content. Moderators moderate by publishing a \emph{moderation stream} associated with a particular content stream that annotates the content. Each user is free to choose the moderator(s) whose streams they wish to subscribe to. Users then apply moderation streams as filters over the raw data of each content stream when they download and view content. Additionally, moderation streams may be further combined, processed, and re-published by other moderators. As with the content streams, the actual implementation could be as simple as a JSON format of allow/deny lists signed by the publisher with a reference to the content stream.

\vspace{1mm}
\noindent
\textbf{Aggregation:} We envision a plethora of user-facing applications to be implemented on top of content and moderation streams to re-aggregate them into coherent user experiences. These aggregators are analogous to, and may superficially appear identical to, the platforms and services that we use today (e.g., Google, Twitter, or Reddit) while implementing completely different content publication and moderation schemes underneath. Alternatively, our design could be implemented is as a content overlay on top of these existing services.

Although we do not elaborate on the specifics of a moderation standard, even a simple moderation model is sufficient to implement the moderation schemes of existing centralized social media platforms.
% like reddit or twitter
For example, to replicate Reddit, an aggregator could serve a forum interface with a fixed set of moderators for each sub-forum whose moderation streams (i.e., deny lists) are unioned together to remove posts and/or users in the forum's data stream. Similarly, to implement a Twitter-like user experience, an aggregator could serve the individual published streams of each user with follows being subscriptions to moderators who publish allow lists. Each user would then only see tweets from the users they follow.

\section{Discussion} \label{sec:discussion}

In networking and distributed systems, decentralizing control and decoupling authority from infrastructure have been studied for decades through various approaches including peer-to-peer systems, homomorphic encryption on the cloud, and blockchain-based systems. Each of these directions focuses on lower level concerns such as storage, data, and communication. In contrast, we identify opportunities to restructure aspects typically considered to be in the ``application layer.'' Unlike many open source and anti-censorship solutions that support savvy activists~\cite{lantern, dingledine2004tor}, we attempt to meet the needs of the uninvolved ‘average’ user.

% decentralized discussion has been around for a long time, why will this work now/better?
Social communication and sharing platforms such as Usenet~\cite{usenet}, one of the oldest messaging platforms on the Internet, offered a decentralized, distributed online forum with no single authoritative entity that owns the network. Internet Relay Chat (IRC)~\cite{oikarinen1993rfc1459} is one of the earliest group-communication, instant-messaging protocols. Through IRC clients, users can connect to different IRC servers (networks) operated by different parties. This is similar to Usenet in that no single centralized entity controls the whole network. Usenet suffers from issues relating to the storage of undesirable content and, along with IRC, faded in popularity with the rise of modern social media platforms. Our design is comparatively more decentralized and flexible in terms of infrastructure and moderation while still allowing content to be presented as a coherent platform experience.

% some evidence of why this could be adopted
Although users ultimately have to decide whether they would prefer decentralized and open platforms, there is already some evidence suggesting that they would be welcomed. One million users have already demonstrated that they are willing to use even "everything goes" platforms like Gab. Over two million users are active on Mastodon~\cite{mastodon}, which follows a federated model where each instance runs its own membership, content, and moderation policies. Mastodon also boasts over 4,000 such instances, which are discoverable via an unofficial directory that is analogous to indexing on an aggregator. Unlike Mastodon or Diaspora~\cite{diaspora} where federated instances are isolated from each other, our design allows content and moderation to be fluidly mixed and combined. Since our design is general enough to support content structures and moderation policies of existing platforms, we believe that our design could provide a narrow waist for a marketplace of platforms~\cite{protocolsnotplatforms}.

%benefits of transparency
Beyond replicating existing platforms, we expect moderation streams to enable an ecosystem of social media platforms that combine the various streams in novel and useful ways. Our design could help address some of the other major issues with online social networks like gatekeeping bias, algorithmic bias, and echo chambers~\cite{socialmediabias, partisansharing, echochambers} because moderation actions in our design are completely transparent. For example, a Reddit clone could adjust the site's interface to allow users to: view moderation streams, see how moderation streams differ from raw data streams, automatically rank moderators based on certain metrics, and compare moderation streams to get a balanced view or highlight points of contention. Thus, our architecture enables more flexibility, transparency, and accountability than today's social media systems.

\subsection{Removing Unacceptable Content}

Where there is free speech, there is undesirable content. Spam could be discouraged through the use of storage fees or off-the-shelf techniques such as spam filters or web of trust. Other undesirable speech (e.g., gore) can be filtered out by creating a moderator stream using the same policy decisions as is done today.
% multiple jurisdictions and views
A more difficult challenge is deciding between what constitutes acceptable or unacceptable speech. This is an inherently thorny problem because judgements are different between people and across communities or cultures; judgements based on legality are also dependent on jurisdiction and therefore similarly problematic. We address this problem by providing a general mechanism to implement arbitrary filtering policies and support a pluralism of views.

% moderating reprehensible/illegal content
Although using moderation streams as filters allow users to avoid seeing unwanted content, the content itself is still stored in the data layer and visible to others. Unlike today's centralized systems where a request can be made to take down content, our architecture is decentralized and content is potentially harder to take down.
Two aspects of our design help address the problem of storing reprehensible or illegal content: 1) storing a content stream does not directly store content; and 2) all parties, including providers of storage and bandwidth, can refuse to store or serve any content or content streams. Moderation streams directly support such refusals.

% child porn argument -- no worse than FB
Deleting illegal content manually is extremely onerous for independent service providers to manage. One way this issue is resolved today is watchdogs like the Internet Watch Foundation (IWF) publish URL lists to block abhorrent content such as child porn~\cite{IWF}. Similarly, `authoritative' real-world entities such as the FBI or IWF could publish signed moderation streams that are applied by default on aggregators and stand-alone clients. Users would still be free to forcibly disable such filters, but they would also be personally liable and prosecuted for illegal activities through existing law enforcement methods. This new arrangement improves the state of affairs by enabling operators to gracefully relinquish authority over online speech, which they appear to be slowly conceding anyway~\cite{bluesky, fbsupremecourt}. This new ecosystem also could help resolve the issue of applying moderation across jurisdictional contexts to both implement policies with international consensus and enforce local regulations.

\subsection{Adoption and Re-centralization}

%incremental adoption
Our design takes a clean slate approach, but we envision an incremental path to adoption or partial adoption for existing platforms. Some existing corporations have expressed interest in opening up their platforms~\cite{bluesky, dtp}, but they are incentivized to continue walling off content to increase profits by, for example, paying content producers or moderators for exclusive rights~\cite{joeroganspotify}. Content and moderation standardization coupled with legislation could provide sufficient support for corporations to unbundle moderation from content in a manner that leads to independent moderators.

%incentives
We do not detail the economic incentives for the various stakeholders in our ecosystem, but since many of the proposed roles have existing analogues we expect compatibility with Internet business models while offering opportunities for greater flexibility. Content creators, moderators, and aggregators are not compensated directly within our design. Instead, we expect these entities to rely on out-of-band compensation for their services. For example, a moderator (i.e., an ‘influencer’) may earn money through tipping, subscriptions, or sponsorship, while syndicating their moderation feed across aggregators. An aggregator could serve advertisements or offer subscriptions while serving content and pay content creators and moderators.

%lot of problems, re-centralization is a big one
While we have attempted in our design to consider some of the obvious and immediate challenges, problems external to the architecture itself are innumerable. Most prominently, preventing re-centralization is difficult and not a purely technical problem.
The original Internet was also decentralized, but still became centralized and balkanized. Just because an architecture is decentralized and `open' does not mean that its deployment and eventual evolution will be democratic. 

Assuming that our architecture were adopted, we expect that a certain degree of centralization would naturally emerge. A small number of aggregators would eventually become the most popular. Given the seemingly inevitable centralization brought on by economies of scale, our underlying goal is not to reject centralization itself, but rather to prevent the authoritarian consequences that centralization often brings about. As such, the design motif we introduced of intentional impermanence in conjunction with policy support could act as an ``escape hatch'' against autocratic lock-in. Users who take issue with one aggregator would have the freedom to seamlessly switch providers while retaining their content and moderation streams. However, our architecture only works if it remains in use, which is where open standards and protocols and legislation would play a role.

\section{Conclusion}
We have proposed a starting point to enable free speech online while enabling user-desired content moderation. Our design is by no means concrete or comprehensive, but we believe that preserving a moderated user experience is essential to any solution and that the key ideas of decoupling moderation from content, decentralizing moderation, and intentional impermanence are promising directions for future research.

\bibliographystyle{abbrv} 
\begin{small}
\bibliography{main}
\end{small}

\end{document}